\documentclass[prd,12pt,aps,superscriptaddress,preprintnumbers,tightenlines,floatfix,nofootinbib,amssymb]{revtex4}
\usepackage{epsfig}

\newcommand{\beqn}{\begin{eqnarray}}
\newcommand{\eeqn}{\end{eqnarray}}
\newcommand{\eq}[1]{(\ref{#1})}
\newcommand{\Tr}{{\mathrm{Tr}}\,}
\newcommand{\dd}{{\mathrm{d}}\,}
\newcommand{\cZ}{{\mathcal Z}}
\newcommand{\logo}{\\ \vskip -20mm \leftline{\includegraphics[scale=0.3,clip=false]{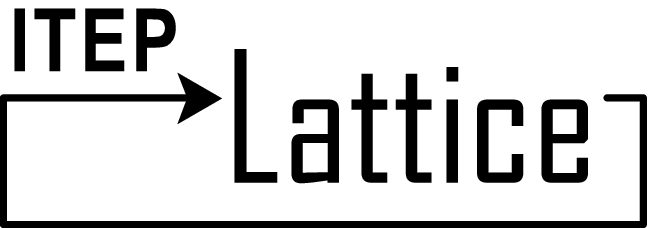}} \vskip 14mm}

\begin{document}

\preprint{ITEP-LAT/2008-12}

\title{Monopoles and vortices in Yang-Mills plasma\,\footnote{
Based on talks at 13th Lomonosov Conference, Moscow, Russia, 23 - 29
August, 2007; and at the conference "Strong Coupling: from Lattice to
AdS/CFT",  Florence, Italy, June 3 - 5, 2008.
}\logo}

\author{M. N. Chernodub}
\affiliation{ITEP, B.Cheremushkinskaya 25, Moscow, 117218, Russia}
\author{V. I. Zakharov}
\affiliation{ITEP, B.Cheremushkinskaya 25, Moscow, 117218, Russia}
\affiliation{INFN -- Sezione di Pisa,  Largo Pontecorvo 3, 56127 Pisa, Italy}

\begin{abstract}
We discuss the role of magnetic degrees of freedom in Yang-Mills plasma
at temperatures above and of order of the critical temperature $T_c$.
While at zero temperature the magnetic degrees of freedom are condensed
and electric degrees of freedom are confined, at the point of the phase transition
both magnetic and electric degrees of freedom are released into the thermal
vacuum. This phenomenon might explain the observed unusual properties of the plasma.
\end{abstract}

\maketitle

\section{Introduction}

At vanishing temperature, the Yang-Mills theories exhibit the
phenomenon of the color confinement. In case of pure Yang-Mills
theories, on which we concentrate below, the criterion of the
confinement is the area law for large enough Wilson loops:
\begin{equation}\label{wilson}
\langle W(C)\rangle~\sim~\exp{\big(-\sigma\cdot
A_{\min}\big)}
\qquad \mbox{or} \qquad
V_{\bar{Q}Q}(R) \to~\sigma\cdot R\,.
\end{equation}
Here $C$ is the rectangular contour with time and space dimensions $T$ and $R$ respectively,
$A_{\min}\equiv R\cdot T$ is the minimal area of a surface spanned on the contour $C$,
$\sigma\neq 0$ is the confining string tension, and $V_{\bar{Q}Q}(R)$ is the heavy-quark
potential at large distances $R$ between quark and antiquark.

While there is yet no understanding of the confinement from the
first principles in the non-Abelian case, it can be modeled in
Abelian theories. In particular, in ordinary  superconductor the
potential between external magnetic monopoles grows linearly at
large distances~$R$,
\beqn
V_{\bar{M}M}~\sim~\sigma_M\cdot R\,,
\eeqn
mimicking
the heavy-quark potential (\ref{wilson}). The microscopical
mechanism behind this example is the condensation of the Cooper pairs.
In the effective-theory language the mechanism is a condensation of
the electrically charged scalar field:
\begin{equation}
\langle \phi_{\mathrm{el}}\rangle~\neq~0\,.
\end{equation}
By analogy with this, well understood case it was speculated long
time ago \cite{DualSuperconductor} that in the non-Abelian theories
it is the condensate of magnetically charged field,
\begin{equation}\label{model}
\langle \phi_{\mathrm{magn}}\rangle~\neq~0\,,
\label{eq:magn}
\end{equation}
that ensures  confinement of  color charges (\ref{wilson}).

Within this general framework of the dual-superconductor model
(\ref{model}) the main question is: what is the microscopical
mechanism behind (\ref{model}). In other words, the question is what
is a Yang-Mills analogue of the Cooper pairs of the ordinary superconductor?
(Remember that we are considering pure Yang-Mills theories without
fundamental scalar fields.)

A clue to the answer to this question might be provided by the
example of the so called compact $U(1)$ theory \cite{ref:compact}
where the magnetic degrees of freedom are identified as topological
excitations of the original theory. In more detail, the
Lagrangian is the same as for a free electromagnectic field:
\begin{equation}
L_{U(1)}~=~{1\over 4 e^2}F_{\mu\nu}^2\,,
\end{equation}
supplemented, however, by the condition that the Dirac string
carries no action. The condition is automatically satisfied in the
lattice, or compact version of the theory.

Since the Dirac string is not seen (it costs no action)
there appear particle-like excitations, or  magnetic monopoles with magnetic
field
\begin{equation}
{\bf H}~\sim~{const\over e}{{\bf r}\over r^3}\,,
\label{eq:H}
\end{equation}
where $e$ is the electric charge and the constant is determined by
the Dirac quantization condition. Equation~\eq{eq:H} corresponds to
the static monopole located at the origin of the coordinate system.

Admitting singular fields, or monopoles into the theory violates
Bianchi identities and modifies the equations of motion:
\begin{eqnarray}
\partial_{\mu}F_{\mu\nu} & = &0\,,
\nonumber\\
\partial_{\mu}\tilde{F}_{\mu\nu} & \equiv & j_{\nu}^{\mathrm{mon}}\,, \label{current}
\\
\partial_{\nu} j_{\nu}^{\mathrm{mon}} & = & 0\,,
\nonumber
\end{eqnarray}
where $j_{\nu}^{\mathrm{mon}}$ is the monopole current.
Moreover, the non-vanishing, conserved current $j_{\nu}^{\mathrm{mon}}$
can be traded for a magnetically charged scalar  field
$\phi_{\mathrm{magn}}$, which appeared previously in Eq.~\eq{eq:magn}.
In order to prove this fact, one uses the so
called polymer representation of field theory in the Euclidean
space-time
(for review of this representation see, {\it e.g.}, Ref.~\cite{ambjorn} while
specific applications to the lattice monopoles are discussed, in particular, in
Refs.~\cite{stone}).

Thus, in case of the compact $U(1)$ theory we have both
microscopical description of the magnetic degrees of freedom in
terms or the topological excitations, or monopoles, and
macroscopical description in terms of the magnetic condensate
(\ref{model}). In the most interesting case of non-Abelian theories
we are not far now from completing a similar program. A specific
feature here is that knowledge on the topological excitations
emerged mainly from the lattice studies. The topological excitations
appear to be the magnetic monopoles and the magnetic vortices. The observed
percolating properties of the monopoles and the vortices are revealing
the nature of the magnetic condensation in the non-Abelian case.

  Reviewing  properties of the topological excitations
  in the Yang-Mills case is a subject of the present notes. Such
  kind of a review could be written, however, a few years ago as well.
  What we are adding, is a discussion of much more recent development
  which is related to the role of the magnetic degrees of freedom
  at temperatures above the temperature $T_c$ of the
  confinement-deconfinement phase transition. Namely, it was
  conjectured in Refs. \cite{ref:PRL,Shuryak1,Shuryak2} that above $T_c$
  the magnetic degrees of freedom constitute an important fraction
  of the Yang-Mills plasma responsible for the unusual properties of
  the plasma which appears to be  rather similar to an ideal liquid
  than to an ideal gas of gluons (for review see \cite{Shuryak4}).

In our presentation, we follow the logic of Ref.~\cite{ref:PRL}. It
is  natural to expect that at the point of the phase transition
the condensate (\ref{model}) is destroyed:
\begin{equation}\label{destruction}
\langle \phi_{\mathrm{magn}}\rangle(T_c)~=~0\,.
\end{equation}
Moreover, it is not a mere speculation but a hypothesis supported by
the lattice data (as we review later). The main question is: which
symmetry is restored when the condensate (\ref{model}) is destroyed?
For example, in case of the Standard Model one is commonly assuming
that at high temperatures all the particles become massless, as in
the original Lagrangian. In our case of non-Abelian field theories
one can speculate that, in some sense,
it is symmetry between electric and
magnetic degrees of freedom which is restored upon the destruction
(\ref{destruction}) of the magnetic condensate by the thermal
fluctuations. Moreover, if this symmetry is indeed restored, then
one should not think about the Yang-Mills plasma as a plasma of
gluons alone but as of plasma with a magnetic component. Thus,
restoration of symmetry could be a general phenomenon behind
observation of unusual properties of the plasma. All these
speculations can be checked, and partially have already been checked
on the lattice.

\section{Phenomenology of Monopoles in Yang-Mills theory}

Historically, the ideas about the nature of  confinement in the
non-Abelian case were strongly influenced by the well-understood
Abelian examples mentioned above. However, in Yang-Mills theories
the effective coupling is large at large distances and  no reliable
analytical tools exist.  As a result, numerical  simulations on the
lattice acquired central role in exploring  the monopole mechanism
of the color confinement.

The first, and apparently serious obstacle on this path is
difficulty to define the monopoles. Indeed, the monopoles are
intrinsically Abelian objects. In particular, in the absence of
scalar fields there are no monopole-like solutions with finite
energy in the non-Abelian case.

A way out was found with the help of
the Abelian projection method~\cite{AbelianProjections}. The basic
idea behind Abelian projections is to fix partially the non--Abelian
gauge symmetry up to an Abelian subgroup. In the physically relevant
$SU(N)$ Yang-Mills theories the residual Abelian subgroup is compact
due to the compactness of the original non-Abelian group. It is the
compact nature of the residual gauge subgroup which leads to
emergence of the Abelian monopoles, see the Introduction.

More specifically, one first utilizes the local gauge invariance to
rotate the non-Abelian gauge field as close as possible
to the Cartan direction in the color space.
Specifically, in the $SU(2)$ gauge theory one minimizes
the lattice analogue of the functional
\beqn
\int {\mathrm d}^4 x \, \Bigl\{[A^1_\mu(x)]^2 + [A^2_\mu(x)]^2\Bigr\}
\eeqn
with respect to the local gauge transformations. This minimization makes
the offdiagonal components small and, thus, this procedure maximizes
the r\^ole of the diagonal (Cartan) elements of the gauge field.
Next, one replaces the original non-Abelian fields by their
projections on the Abelian direction:
\begin{equation}\label{projection}
A_\mu (x) \equiv t^a A_{\mu}^a(x)~\to~ t^3 A_{\mu}^3(x)\,,
\end{equation}
(here $t^a$ are generators of the gauge group),
and then one defines the monopoles in terms of the projected gauge fields
$A_{\mu}^3(x)$ as if they were the fields of a compact $U(1)$
theory. This step is very crucial.

Analytically, it is impossible to evaluate the effect of the
projection (\ref{projection}). However, the  numerical simulations
of non-Abelian gauge theories on the lattice provide us with a
powerful  method to probe the dual superconductor hypothesis. This
``experimental'' method allows us to investigate features of the
Abelian monopoles numerically from the first principles of the
theory. In numerical simulations the formation of the chromoelectric
string can be observed in a very straightforward way. The
contribution of the monopole-like gluonic configurations to the
energy of the string -- probed by quantum averages of time-oriented
Wilson loops -- turns out to dominate over the contribution from
other (``irrelevant'') fluctuations of the gluonic fields. This
property of the confining configurations -- which is also known as
the ``monopole dominance'' -- was clearly demonstrated in
Ref.~\cite{ref:Bornyakov} following the original study of
Ref.~\cite{AbelianDominance}. The monopole dominance confirms that
the color confinement in Yang-Mills
theory originates from a specific dynamics of the Abelian monopoles.

In addition to the monopole dominance one can also observe numerically that the Abelian monopole
currents circulate around the confining chromoelectric flux tube in the same way as the Cooper pairs in
the ordinary superconductor circulate around the Abrikosov tube~\cite{ref:Haymaker}. This is yet another fact
in favor of the validity of the dual superconductor mechanism.
The important cornerstone of the dual superconductivity,
the monopole condensation, was numerically observed in Refs.~\cite{MonopoleCondensation} using
various types of the monopole creation operators.

Monopoles of the maximal Abelian projection possess remarkable
properties which, in particular, allow to visualize formation of the
condensate (\ref{model}).  The emergence of the monopole condensate
at low temperatures can be seen as a percolation of the monopole
world-lines. The percolation  property means that there is a nonzero
probability for two arbitrary infinitely separated points to be
connected by a single monopole trajectory. Thus, a portion of the
monopoles in the confinement phase propagate for the infinitely long
distances forming an essentially infrared (zero-momentum) component
of the monopole density. One can intuitively draw a parallel between
the finite monopole density at exactly zero value of the momentum
and the condensed portion of the Bose-particles which trigger
emergence of either superconductivity (in the case of charged
particles) or superfluidity (in a system of neutral particles).

In Yang--Mills theory the monopole condensate exist at low
temperatures, and the condensate disappears when the temperature reaches its critical value $T=T_c$.
The quarks are liberated in the high temperature phase since the monopoles are not condensed
and the dual Meissner effect is absent at $T>T_c$.
This does not mean, however, that the Abelian monopoles do not play essential role in non-perturbative
physics at high temperature. It is known that in the deconfinement phase the vacuum is made of the ``static''
monopoles which run dominantly along the imaginary time, or ``temperature'', direction in the Euclidean space-time.
The monopoles running in the spatial directions are suppressed. The static nature of the monopoles
is a direct consequence of the dimensional reduction: Fourier components of the gluonic fields with
non-zero Matsubara frequencies have low statistical weight in comparison with the static gluonic fields
characterized by the zero Matsubara frequency. The static nature of the monopoles agrees well with the absence
of confinement at high temperatures. According to numerical simulations~\cite{AbelianDominanceT,Ejiri}, in the
deconfinement phase the static monopoles provide a dominant contribution to the so called ``spatial string
tension'' which corresponds to a coefficient in front of the area term in the quantum average of sufficiently
large spatial Wilson loops. Thus, the presence of the monopole-dominated spatial string tension is the first
important indication that the monopoles may play a nontrivial role not only in the low temperatures but also
in the high temperature phase.

\begin{figure}[!htb]
\begin{center}
\includegraphics[angle=-0,scale=0.75,clip=true]{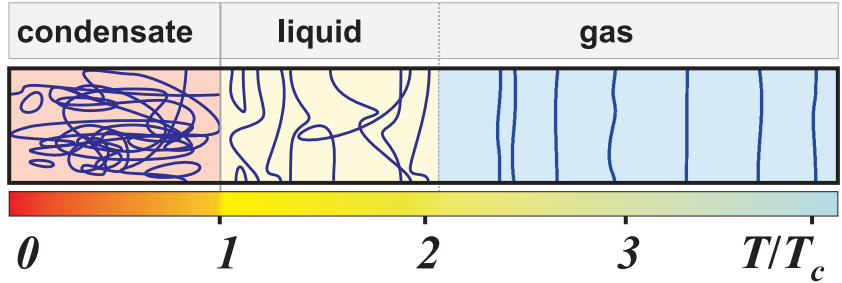}
\end{center}
\caption{Illustration of the behavior of the monopole trajectories in Yang-Mills theory.}
\label{fig:states}
\end{figure}
An illustration of the behavior of the monopole trajectories in the finite-temperature Yang--Mills theory is
provided in Figure~\ref{fig:states}. As we have already mentioned, at low temperature the monopoles form a condensed state
which means that there is a long (``infrared'') monopole trajectory which fills the whole space. Contrary to the
perturbative monopoles, which are represented by short (``ultraviolet'') trajectories, the condensate
is of a non-perturbative nature. In the deconfinement phase the condensate is broken and the infinitely
long (in the spatial dimensions) trajectory does not exist. The condensate melts into a monopole liquid~\cite{ref:PRL}
which is evaporated into a monopole gas at very high temperatures. The intermediate monopole liquid state was also
predicted in Refs.~\cite{Shuryak1,Shuryak2}
while the formation of the monopole gas at very high temperatures was discussed in
Refs.~\cite{chris,ref:blocking} (see also references therein).
In the liquid state the monopoles become specifically correlated as one can
observe from the accurate lattice data of Ref.~\cite{ref:Italians}. In Figure~\ref{fig:states} the high temperature
gas of the Euclidean monopoles is seen as a collection of the static worldlines. Below we discuss the phase diagram
of Figure~\ref{fig:states} is more details.

\section{(Abelian) monopoles and (Center) vortices}

The Abelian monopoles in Yang-Mills theory are not emerging alone in the vacuum. The monopoles are known to
be related to other objects of higher dimensionality which are known as center vortices. Historically, the center
strings were invented as an alternative to the monopole-based mechanism of the quark confinement~\cite{ref:center}.
In the vortex--based picture the quark confinement emerges due to spatial percolation of the vortices
which lead to certain amount of disorder. Each vortex piercing the Wilson loop changes the
value of the Wilson loop by a center element. Thus, sufficiently large loops receive wildly fluctuating
contributions from the vortex ensembles making the average value of the Wilson loop very small. One can show that
the suppression of the loop follows the area law for large loops~\cite{greensite}.

Later it turns out that both the dual superconductor and the monopole-based mechanisms of the confinement
are complimentary to each other being a reflection of a genuine property of the confining non-Abelian vacuum.
The Abelian monopoles and the center vortices appears to be strongly correlated
with each other~\cite{ref:chains1,ref:chains2}: almost all monopoles are sitting on top of the vortices.
Alternatively, on can say that the center vortices are passing thought Abelian monopoles.
The fact of the correlation indicates that the Abelian monopoles and the
center vortices are constituents of a generic gluonic object in which the two neighbor
monopoles are connected together by a segment of the vortex.
In $SU(2)$ gauge theory this object is considered as a monopole-vortex chain, while in the
$SU(3)$ case the objects form the monopole-vortex 3-nets.
The formation of the chains and nets is essential for the self--consistent
treatment of the monopoles in the quark-gluon plasma~\cite{ref:PRL}.

In $SU(2)$ gauge theory the center vortex can be regarded as an Abelian vortex carrying
the flux which is equal to a half flux of the monopoles. Indeed, the distribution of the
magnetic part of the gluon energy density is not spherical as one could guess from a na\"ive
Coulomb-like distribution of the magnetic fields around the monopole~\cite{ref:chains2}.
Each monopole is a source of two vortex fluxes which must be connected to other anti-monopole(s)
due to conservation of the vortex flux. As a result, there appears
either monopole-antimonopole pairs or the closed set of the vortex segments which
connect alternating monopoles and antimonopoles. It is clear that the described vortex chains
are not globally oriented similarly to the center vortices.

Similar monopole-vortex chains were found in numerous (non-)supersymmetric non-Abelian gauge theories
involving various Higgs fields~\cite{ref:monvort}.

\begin{figure}[!htb]
\begin{center}
\includegraphics[angle=-0,scale=0.75,clip=true]{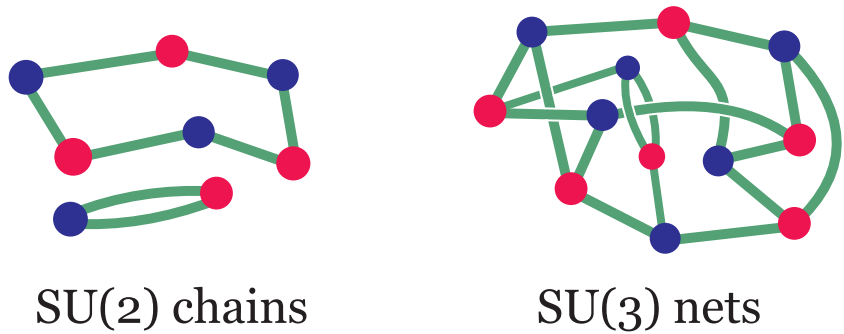}
\end{center}
\caption{The monopole-vortex chains and the monopole-vortex nets in $SU(2)$ and $SU(3)$ gauge theories, respectively.}
\label{fig:chains}
\end{figure}

A bit more complicated situation appears in $SU(3)$ gauge theories. In this case the vortex
carry one third of the total monopole flux and thus the three vortices may meet at a single
point. The fluxes of the vortices should add to zero (modulo 3).
Thus, in $SU(3)$ gauge theory the monopoles and vortices form 3-nets instead of the chains.
The monopole-vortex junctions are considered in details in Ref.~\cite{ref:nexus}, where they
are called as ``nexuses''. In general case of $SU(N)$ gauge group each nexus is the source
of $N$ vortex pieces. We visualize examples of the $SU(2)$ monopole-vortex chains and
their $SU(3)$ net-like counterparts in Figure~\ref{fig:chains}.

Both the vortex chains and the $N$--nets has Abelian analogues in the Abelian gauge theory with compact gauge field
and the double or, respectively,
$n$--charged Abelian Higgs field. The monopole-vortex structures in such theories can easily
be identified in computer simulations, see, for example, Ref.~\cite{ref:Schiller} for the double charged gauge field, $n=2$.
It is possible to argue that in the Abelian gauge theories the monopoles and vortices must be geometrically related
to each other in order to ensure the stability of the confining string stretched between the fundamentally charged ($q = \pm e$) test particles,
and, simultaneously, to guarantee a breaking of the string spanned between certain multiply charged (i.e, $q = \pm n e$) particles,
Ref.~\cite{ref:Schiller}. Following this logic we expect that in the $SU(2)$ gauge theory without matter fields the monopoles
must form monopole-vortex chains to ensure stability (string breaking) of the chromoelectric string in the fundamental (adjoint)
representation. The same logic inevitably uncovers existence of the monopole-vortex nets in pure $SU(3)$ Yang-Mills theory.

\section{Thermodynamics of Yang-Mills theories \newline and topological objects}

The thermodynamical properties of Yang-Mills theory are of particular interest
in view of ongoing and planned experiments on heavy-ion collisions. In these
experiments the heavy ions collide forming a fireball with thermodynamically
large number of particles. In these collisions the thermodynamics of Yang-Mills
fields become relevant. Thus, it is important to know if the monopoles and
vortices -- which are understood as singular gluonic configurations --
contribute to the thermodynamical properties of the plasma or not?

Basic thermodynamical quantities of Yang-Mills theory,
such as the pressure $p$ and the energy density $\varepsilon$, respectively,
\beqn
p(T) = T^4 \int\limits^T \ \frac{{\mathrm{d}}\, T_1}{T_1} \ \frac{\theta(T_1)}{T_1^4}\,, \qquad
\varepsilon(T) = 3 \, T^4 \int\limits^T \ \frac{{\mathrm{d}}\, T_1}{T_1} \ \frac{\theta(T_1)}{T_1^4} + \theta(T)\,,
\label{eq:pressure:anomaly}
\eeqn
can be determined from the anomalous quantum average of the trace
\beqn
\theta(T) = \langle T^\mu_\mu \rangle \equiv \varepsilon - 3 p\,,
\label{eq:anomaly:continuum}
\eeqn
of the energy--momentum tensor,
\beqn
T_{\mu\nu} = 2 \, \Tr \left[G_{\mu\sigma} G_{\nu\sigma} - \frac{1}{4} \delta_{\mu\nu} G_{\sigma\rho} G_{\sigma\rho}\right]\,.
\eeqn
Here $G_{\mu\nu} = G_{\mu\nu}^a t^a$ is the field strength tensor of the gluon fields $A_\mu$.

The {\it bare} Yang--Mills theory is a conformal theory and therefore at the classical level the energy--momentum tensor
is traceless. However, because of a dimensional transmutation the conformal invariance is broken at the quantum level
and the energy--momentum tensor exhibits a trace anomaly which is expressed via the expectation value of the action
density~$\frac{1}{2} \Tr G_{\mu\nu}^2(x)$,
\beqn
\theta = \left\langle \tilde\beta(g)\Tr G_{\mu\nu}^2(x) \right\rangle\,,
\eeqn
This definition involves the $\beta$-function of Yang-Mills theory:
\beqn
\tilde\beta(g) \equiv \frac{\beta(g)}{g}  =
\frac{{d} \log g}{{d} \log \mu} = - g^2 (b_0 + b_1 g^2 + \dots)\,.
\label{eq:theta}
\eeqn
The trace anomalies~\eq{eq:anomaly:continuum} and the thermodynamical quantities~\eq{eq:pressure:anomaly}
were calculated in Refs.~\cite{Engels:1988ph} and \cite{Boyd:1996bx} for $SU(2)$ and $SU(3)$ gauge theories,
respectively. In Eq.~\eq{eq:theta} the finite-temperature expectation value of the action density should be
normalized by subtraction of the corresponding $T=0$ contribution.

Relevance of the magnetic monopoles to the thermodynamics of Yang--Mills theory can already be guessed
from particular results obtained at zero temperature. Indeed, the monopoles appear to
carry an excess of the (magnetic part of) non-Abelian action density~\cite{ref:physical} which at
non-zero temperature contribute to the trace anomaly~\eq{eq:theta} and, consequently, to the
pressure and energy density of the Yang-Mills fields~\eq{eq:pressure:anomaly}.
Note that the infrared and ultraviolet monopoles provide different contributions to the trace anomaly:
the ultraviolet monopoles carry an excess of the non-Abelian action while the
infrared monopoles are deficient in the density of the non-Abelian action~\cite{ref:physical:next}.

It was observed in numerical simulations of Ref.~\cite{ref:physical} that the normalized correlator between
the monopole trajectory $j_\mu(x)$ and the action density,
\beqn
C = \frac{\langle \frac{1}{2} \Tr {[j_\mu(x) {\widetilde G}_{\mu\nu}(x)]}^2 \rangle}{
\langle j^2_\alpha(x)\rangle \,
\langle \frac{1}{2} \Tr {G_{\beta\gamma}^2(x)} \rangle} - 1\,,
\label{eq:correlation}
\eeqn
is always positive at zero temperature
(we treat both the infrared and the ultraviolet monopole trajectories on equal footing in this calculation).
The magnetic nature of the correlation~\eq{eq:correlation} is stressed by the dual field strength tensor,
\beqn
{\widetilde G}_{\mu\nu} (x) = \frac{1}{2} \varepsilon_{\mu\nu\alpha\beta} G_{\alpha\beta}(x)\,.
\eeqn
If the monopole is static, $j_\mu \sim \delta_{\mu 4}$ then
only spacelike (magnetic) components of the field strength tensor $G_{ij}$, $i,j=1,2,3$ contribute
to the correlator~\eq{eq:correlation}.

Another important consequence of the non-zero correlation~\eq{eq:correlation} become obvious
if one remembers that the magnetic monopoles in Yang-Mills fields are identified with the
help of the Abelian projection method~\cite{AbelianProjections} which involves a particular
gauge fixing. The fact of the nonzero correlation between the ("gauge-fixed") Abelian monopoles and the gauge-invariant
action density of the non-Abelian fields shows that the monopoles are not just the artifacts
of the gauge fixing, as they are able to pinpoint magnetically energetic parts of the non-Abelian vacuum.

Similarly to the Abelian monopoles~\cite{ref:physical}, the center vortices also exhibit a local correlation with
the action density~\cite{ref:chains2} which can be probed by a correlator which is analogous to Eq.~\eq{eq:correlation}.
This fact is not unexpected because the monopoles and vortices are interrelated with each other
and thus the vortices can also contribute to the thermodynamics of the system as well.

Dedicated numerical simulations in $SU(2)$ and $SU(3)$ gauge theories confirm that both the monopole--vortex chains
in $SU(2)$ gauge theory and their $SU(3)$ counterparts, the monopole--vortex nets, are thermodynamically relevant
degrees of freedom in the gluonic plasma~\cite{ref:EOS} as both these topological degrees of freedom do contribute
to the trace anomaly. The contribution of the monopoles is positive while the vortices provide a large negative
contribution. The negative value of the vortex contribution to the anomaly is in agreement with
general theoretical expectations~\cite{Gorsky:2007bi}.

\section{Properties of monopoles at finite temperature}

It was suggested in Refs.~\cite{chris,ref:PRL,Shuryak1} that there
is a magnetic component of Yang-Mills plasma which is crucial for
determining properties of the gluon plasma realized in the
Yang-Mills vacuum above the critical temperature.

\begin{figure}[!htb]
\begin{center}
\includegraphics[angle=-0,scale=0.75,clip=true]{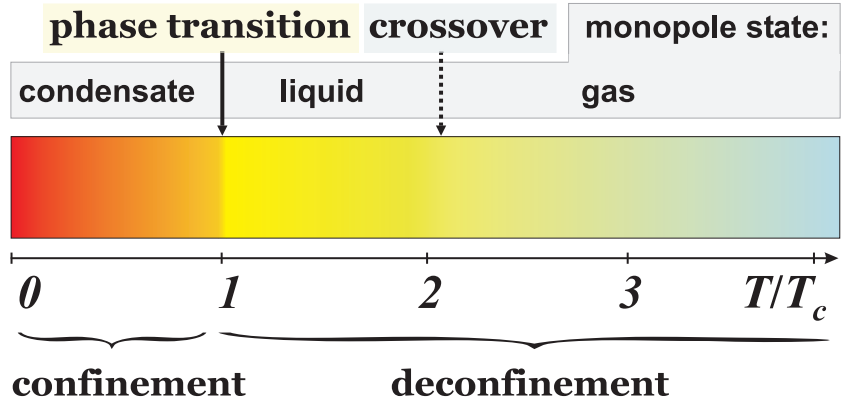}
\end{center}
\caption{States of the Abelian monopoles in Yang-Mills theory, suggested in Ref.~\cite{ref:PRL}.
The thermodynamic (phase) transitions and phases of the gluonic vacuum are also indicated (a detailed description is given in the text).}
\label{fig:scales}
\end{figure}

In Ref.~\cite{chris,Shuryak1} the constituents of the magnetic
component are thought to be classical magnetic monopoles. In
Ref.~\cite{ref:PRL} the magnetic component is identified with the
magnetic strings which join (nonclassical) monopoles constituting
chainlike and netlike structures. The suggested phase diagram of the
topological magnetic component of the Yang-Mills theory is presented
in Figure~\ref{fig:scales}.
\begin{itemize}
\item[$\blacktriangleright$]
At low temperatures, $T < T_c$, the monopoles form a condensate which determines the dual superconducting
properties of the Yang--Mills vacuum. The monopole condensate is responsible for the confinement of color.
\item[$\blacktriangleright$] The phase transition, $T = T_c$, marks complete melting of the condensate and
appearance of the magnetic monopoles in a liquid state in analogy with a bosonic system such as
helium ${}^{4}$He. In Ref.~\cite{ref:PRL} the liquid state is assumed to be realized in the range of temperatures
$T_c < T \lesssim 2 T_c$. In the pure liquid monopole state the confining properties of the theory are lost.
Note that the monopole liquid state may coexist with the monopole condensate even {\it below} $T_c$.
\item[$\blacktriangleright$] At even higher temperatures the monopoles are suggested to be in a gaseous state.
The liquid state and the gaseous state are separated by a crossover which is an analytic (smooth) transition from
one phase to the other. Due to the crossover the difference between the liquid and the gas regions is somewhat blurry
and the liquid state contains an admixture of the gas while the gas state possess a fraction of the liquid.
In Ref.~\cite{Shuryak2} it was argued that the liquid state of monopoles survives at much higher temperatures as $2 T_c$.
\end{itemize}

A direct check of the condensate-liquid-gas transition can be done numerically with the help of the
lattice simulations of Euclidean Yang-Mills theory. At finite temperature one of the Euclidean directions
is compactified to form a circle of the length which is equal to the inverse temperature.
The lattice simulations provide us with thermodynamical ensemble of the configurations of non-Abelian gauge
field. These configurations can be used to determine the worldlines of the Abelian monopoles as well as
the worldsheets of the Center strings corresponding to the gauge theory in thermal equilibrium.
However, the ensembles of the monopole trajectories should contain both the virtual monopoles belonging to
the vacuum and, at the same time, the thermodynamically relevant (or, real) monopoles. In order to measure
physically meaningful observables in continuum limit, one should be able to separate these types of monopoles
from each other.

\begin{figure}[!htb]
\vskip 5mm
\begin{center}
\includegraphics[angle=-0,scale=0.55,clip=true]{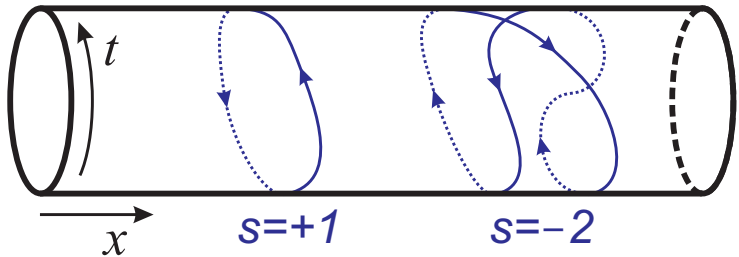}
\end{center}
\caption{An illustration of the thermal (real) monopole trajectories in the Euclidean space-time.
These trajectories correspond to one monopole (with the winding number $s=+1$) and to two antimonopoles
represented by a single trajectory with the winding number $s=-2$.}
\label{fig:winding}
\end{figure}

According to Ref.~\cite{ref:PRL} the thermodynamically relevant monopoles can be distinguished
from the virtual monopoles by a simple principle: the thermal monopoles should wrap around the
temperature (compactified) direction of the Euclidean space. Moreover, the quantum density of
the thermal monopoles is not equal to the average density of the total length of the (wrapped)
monopole trajectories as one could na\"ively guess on general grounds. Such a quantity is
divergent in the ultraviolet regime making its interpretation somewhat obscure. The density of
the thermal monopoles corresponds to the density of the winding number $s$ of the monopole trajectories
in the temperature direction of the Euclidean space:
\beqn
\rho = \frac{1}{V_{3d}} \langle |s| \rangle\,.
\label{exactt}
\eeqn
Such trajectories are illustrated in Figure~\ref{fig:winding}.

These and related quantities were previously measured in Ref.~\cite{Ejiri,ref:MMP}. A thorough calculation of the
winding number in Ref.~\cite{ref:Italians} demonstrated a very good scaling of this quantity towards
the continuum limit.
Moreover, the specific form of the correlations of the (wrapped) monopole trajectories, $\langle j_\mu(x) j_\nu(y) \rangle$
observed numerically in Ref.~\cite{ref:Italians}, provided a strong evidence that the monopoles in
the deconfinement phase form a liquid state~\cite{Shuryak2}.

In the confinement phase at low temperatures the monopoles should also be thermally excited. In other
words, the monopoles should form a pure condensate (plus an admixture of ultraviolet component) in at
any nonzero temperature. Intuitively this fact is clear from an analogy with superfluid Helium-4 which
was put forward in Ref.~\cite{ref:PRL}. In the low temperature phase below the superfluid phase transition
($T_c \approx 2.15$\,K) the thermal (noncondensed) component constitute a sizeable amount of the liquid Helium.
For example, at $1\,{\mathrm{K}} \approx 0.5\, T_c$ less than $10\%$ of the Helium atoms are in the
condensed (zero-momentum) state~\cite{ref:He}.

The direct observation of the wrapped trajectories in numerical simulations of
the lattice Yang-Mills theory would be a difficult task since -- as one can see from
the illustration provided in Figure~\ref{fig:scales} -- the wrapped trajectories will
geometrically link with both with trajectories of the ultraviolet monopoles and with the condensed
infrared components. In order to separate the wrapped (thermal) and spatial (condensate and ultraviolet)
parts of the monopole trajectories one needs to go to the very fine lattices with very small lattice spacing.

At very high temperatures the monopole liquid should evaporate into the gaseous state and eventually disappear at all.
Indeed, a careful accounting for finite-size effects in lattice simulations~\cite{ref:Fodor,ref:Karsch} shows that
at high temperatures the equation of state of Yang--Mills theory converges to the expected Stefan-Boltzmann
limit
\beqn
\varepsilon_{\mathrm{free}}(T)  = 3 P_{\mathrm{free}}(T) = N_{d.f.} \, \frac{\pi^2}{30} \, T^4\,,
\label{eq:SB}
\eeqn
which describes $N_{d.f.} = 2 (N_c^2 - 1) $ degrees of freedom corresponding to
$N_c^2 - 1$ noninteracting gluons with two transverse polarizations. There is no room
for additional degrees of freedom such as monopoles and/or center strings as, if they exist,
they would contribute to the number of degrees of freedom~$N_{d.f.}$ and modify the
Stefan-Boltzmann limit \eq{eq:SB}.

The disappearance of the monopoles at very large temperatures agrees nicely with dimensional
reduction arguments~\cite{ref:Appelquist}. Indeed, the monopoles survive in the high-temperature limit as static
particles, see Figure~\ref{fig:states}. Their properties -- including the magnetic correlation
length -- should be determined by the three-dimensional coupling constant,
\beqn
g^2_{3d} = g^2_{4d} \, T\,.
\eeqn
Thus we expect that the monopole density is proportional to~\cite{chris,ref:PRL}
\beqn\label{density2}
\rho(T) = C_\rho \, g^6_{3d}(T) \propto {\left(\frac{T}{\log T/\Lambda}\right)}^3 \qquad T \gg T_c\,,
\eeqn
where $C_\rho$ is a temperature-independent parameter and $\Lambda \sim \Lambda_{\mathrm{QCD}}$ is a dimensional parameter.
The temperature dependence exhibited by Eq.~(\ref{density2}) can be reproduced by~Eq. (\ref{exactt}) provided
that there exists temperature-dependent chemical potential applied both to the monopoles and to the antimonopoles~\cite{ref:PRL},
\beqn
\mu \sim T \log g^{-6}_{4d}(T) \sim 3 T \log \log T / \Lambda\,.
\label{mu}
\eeqn
The chemical potential suppresses the monopole density at high temperatures
by the logarithmic factor,
\beqn
\exp\{- \mu /T\} \sim g^6_{4d}(T) \sim 1/ \log^3 (T/\Lambda)\,.
\eeqn

In Ref.~\cite{ref:PRL} we stressed that in order to explain the observed magnetic screening the monopole gas should
be treated as strongly interacting even at the asymptotically high temperatures
despite the logarithmic suppression~\eq{density2} of the monopole density.

Thus, the monopoles and the antimonopoles should disappear at very large temperatures. At finite temperatures
the monopoles may form both the gas and the liquid which appear most likely in a mixed state
due to a crossover nature of the analytical transition which separates the gaseous phase from the liquid phase.
However, we may discuss what state, either the liquid or the gas, is dominating at each particular temperature?
A direct answer to this question can be done by investigation of the transport properties of the monopoles
reflected in their mutual correlations.
However, an inspiring hint may also be obtained by analyzing the contribution of the monopoles to the
spatial string tension. This quantity is also sensitive to the inter-monopole correlations.

Suppose that the monopoles are in a purely gaseous state and that they are interacting by a Coulomb law.
Since the monopole trajectories are getting static in the deconfinement phase (see Figure~\ref{fig:states}),
the monopoles can be treated as point--like objects in three space dimensions. Their trajectories
are characterized by a position, $x$, and the magnetic charge, $m$
(defined in units of a fundamental magnetic charge, $g_M$). The simplest model
possessing the gas of the monopoles is the 3D compact quantum electrodynamics (cQED${}_3$) in which the monopole
action is given by the $3D$ Coulomb gas model~\cite{Polyakov}:
\beqn
\cZ = \sum\limits_{N=0}^\infty \frac{\zeta^N}{N!}
\Biggl[\prod\limits^N_{a=1} \int \dd^3 x^{(a)} \sum\limits_{m_a = \pm 1}\Biggr]
\exp\Bigl\{ - \frac{g^2_M}{2} \sum\limits_{\stackrel{a,b=1}{a \neq b}}^N
m_a m_b \, D(x^{(a)}-x^{(b)})\Bigr\}\,.
\label{CoulombModel}
\label{Z1}
\eeqn
The Coulomb interaction in Eq.\eq{Z1} is represented by the inverse
Laplacian~$D$:
\beqn
- \partial^2_i D(x) = \delta^{(3)}(x)\,,
\eeqn
and the latin indices $a,b$ label different monopoles. To get analytical expressions
below we make a standard assumption that the density of the monopoles is low.
The monopole charges therefore are restricted by the condition $|q_a| \leq 1$
which means that the monopoles do not overlap. The average monopole
density $\rho$ is controlled by the fugacity parameter $\zeta$,
giving $\rho = 2 \zeta$ in the leading order of the dilute gas
approximation~\cite{Polyakov}.

The interesting dimensionless quantity which characterizes the thermal mo\-no\-po\-les~is
\beqn
R_{\mathrm{sp}} = \frac{\sigma_{\mathrm{sp}(T)}}{\lambda_D(T)\rho(T)}\,,
\label{Rsp}
\eeqn
where $\sigma_{\mathrm{sp}}(T)$ is the spatial string tension and
$\rho(T)$ is the density of the monopoles.
The magnetic screening length $\lambda_D(T)$ is a non-perturbative quantity
describing a screening of the magnetic gluons in the Yang-Mills plasma.

If the monopoles are in the gas state then from the partition function~\eq{CoulombModel} it follows that the ratio~\eq{Rsp}
should satisfy the following relation:
\beqn
R^{\mathrm{gas}}_{\mathrm{sp}} = 8\,.
\label{eq:theor}
\eeqn
The quantity~\eq{Rsp} was numerically calculated in $SU(2)$ lattice gauge theory using two different approaches~\cite{ref:blocking}
which provided us with similar results which coincide within error bars.
The numerical procedure involved the procedure of the  ``topological blocking from continuum'' which
is omits the contribution of the ultraviolet artifacts leaving us with thermally relevant wrapped trajectories only.
The blocking procedure provided us with the product $\lambda_D(T)\rho(T)$ while the spatial string tension
$\sigma_{\mathrm{sp}}(T)$ was taken from an independent numerical calculations of hot Yang-Mills theory.

\begin{figure}[!htb]
\begin{center}
\includegraphics[angle=-0,scale=0.75,clip=true]{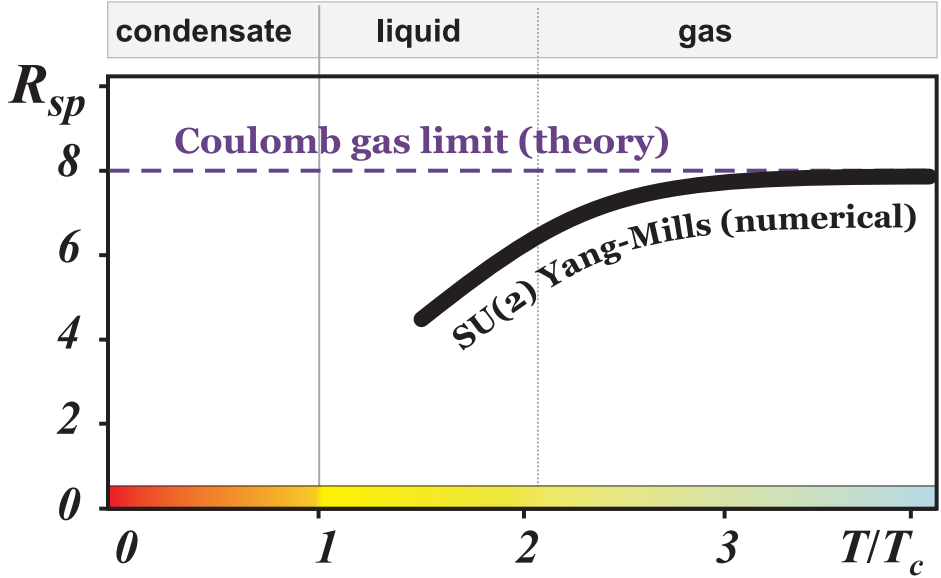}
\end{center}
\caption{A representation of the ratio~\eq{Rsp} calculated in Ref.~\cite{ref:blocking}.
At temperatures $T \gtrsim 2.5 T_c$ the monopoles are dominantly in the gaseous state
as the gas limit \eq{eq:theor} is reached.}
\label{fig:rsp}
\end{figure}

The ratio~\eq{Rsp} is schematically represented in Figure~\ref{fig:rsp}. At relatively low temperatures, $T \lesssim 2.5 T_c$
the numerically calculated quantity deviates substantially from the gas limit. However, as the temperature increases
the gas limit~\eq{eq:theor} is gradually reached. From Figure~\ref{fig:rsp} one can conclude that at
$T \gtrsim 2.5 T_c$ the monopoles are dominantly in the gaseous state.

\section{Conclusion}

 The idea that thermal Yang-Mills plasma contains both electric and
 magnetic degrees of freedom proves promising to understand dynamics
 of the plasma. For the first time, and in a natural way one comes
to reproduce liquid-like properties of the plasma as related to the
magnetic component \cite{ref:PRL,Shuryak1}. It is crucial that
lattice measurements provide with numerically exact answers to some
questions concerning the magnetic component of the plasma. In
particular, one measures directly contribution of the magnetic
degrees of freedom to the free energy and to the density of the magnetic
monopoles as function of the temperature.

On the other hand, further systematic development of the emerging
picture does not look straightforward. One of the problems is that
seemingly both electric and magnetic degrees of freedom are
important and they interact strongly. On the other, direct lattice
measurements are possible now only on the magnetic component.

Also, all the lattice measurements refer to the Euclidean space-time
and their interpretation in terms of the Minkowski space is not
necessarily easy. As a result we do not have yet a clear picture for
the magnetic component of the plasma in conventional terms.

Further lattice measurements seem to be crucial to clarify these
issues. In particular, till now thermal properties of the wrapped
monopole trajectories turned to provide most important information
on the magnetic component of the plasma. Theory of vortices is much
less developed compared to the theory of monopoles since the string
theory in general is much less understood compared to the particle
models.

In conclusion, let us mention an appealing possibility that the
physics of the magnetic strings and the particle physics merge at temperatures
around $T_c$. Indeed, the magnetic strings are the one-dimensional linelike objects
in 3d timeslices of the 4d spacetime. Despite at $T>T_c$ the magnetic strings are
aligned with the time direction they still percolate in the spatial directions.
This means in turn that the one-dimensional imprints of the magnetic strings can be
viewed as percolating trajectories in the 3d timeslices. Since the random trajectories
in space of any dimension correspond to particles, we conclude that the light degrees
of freedom associated with the vortices at $T>T_c$ are particles living in the three
dimensional space.

It was recognized recently \cite{forcrand} that the physics of the
Yang-Mills plasma at $T>T_c$ and close to $T_c$ can be described in
terms of a simple super-renormalizable 3d model if one postulates
existence of scalar particles living in 3d space.

Thus, we are invited to speculate that the 1d defects --
which are projections of the center strings onto the 3d space --
are representing indeed these scalar particles. If so, many properties
of such scalar particles could be clarified through lattice measurements
on the particle trajectories.

\section*{Acknowledgments}
This work was partly supported by the STINT Institutional grant IG2004-2 025,
and by the grants RFBR 06-02-04010-NNIO-a, RFBR 08-02-00661-a, DFG-RFBR 436 RUS,
by a grant for scientific schools NSh-679.2008.2, by the Federal Program of the
Russian Ministry of Industry, Science and Technology No. 40.052.1.1.1112
and by the Russian Federal Agency for Nuclear Power.

\end{document}